\documentstyle[11pt,aaspp,flushrt]{article}	
\def\etal{{\it et al.}}
\def\ie{{\it i.e.}\ }
\def\eg{{\it e.g.}\ }

\begin{document}

\title{The Orbital and Absolute Magnitude Distributions 
                of Main Belt Asteroids.}

\author{R. Jedicke and T. S. Metcalfe\altaffilmark{1}}

\affil{Lunar and Planetary Laboratory, \\
       University of Arizona, Tucson, AZ, 85721 \\
       jedicke@pirl.lpl.arizona.edu \\
       metcalfe@rm-209.lpl.arizona.edu}

\altaffiltext{1}{Present Address:  Department of Astronomy, University of Texas-Austin.}

\begin{abstract}

We have developed a model-independent analytical method for debiasing the four-dimensional $(a,e,i,H)$ distribution obtained in any asteroid observation program and have applied the technique to results obtained with the 0.9m Spacewatch Telescope. From 1992 to 1995 Spacewatch observed $\sim 3740$ deg$^2$ near the ecliptic and made observations of more than 60,000 asteroids to a limiting magnitude of V$\sim$21.  The debiased semi-major axis and inclination distributions of Main Belt asteroids in this sample with $11.5 \le H < 16$ match the distributions of the known asteroids with $H<11.5$.  The absolute magnitude distribution was studied in the range $8<H<17.5$.  We have found that the set of known asteroids is complete to about absolute magnitudes 12.75, 12.25 and 11.25 in the inner, middle and outer regions of the belt respectively.  The number distribution as a function of absolute magnitude can not be represented by a single power-law ($10^{\alpha H}$) in any region.  We were able to define broad ranges in $H$ in each part of the belt where $\alpha$ was nearly constant.  Within these ranges of $H$ the slope does not correspond to the value of 0.5 expected for an equilibrium cascade in self-similar collisions (Dohnanyi, 1971).   The value of $\alpha$ varies with absolute magnitude and shows a `kink' in all regions of the belt for $H\sim13$.  This absolute magnitude corresponds to a diameter ranging from about 8.5 to 12.5 km depending on the albedo or region of the belt.

\end{abstract}

\section{Introduction}

The size distribution for a set of asteroids is crucial to an understanding of their past and continuing collisional evolution.  It has long been recognized that the dips, bumps and slope of the size distribution provide a glimpse into the interaction of objects orders of magnitude larger than any possible terrestrial experiment.  From a high resolution size distribution of Main Belt asteroids it may be possible to deduce the production rate of meteoroids and Near Earth Asteroids, the cratering rate on the inner planets, asteroidal dust production, risk to interplanetary spacecraft, the age of the solar system, the `original' mass in the Main Belt and the impact-strengths of the asteroids.  The problem with obtaining this planetary grail is that the size distribution is not directly observable.

The size distribution of asteroids may be obtained indirectly through a variety of techniques.  An examination of the cratering record on the moon and other objects (\eg Belton \etal\space 1992) has provided an estimate of the size distribution, but the relationship between crater diameter and the size of the incident asteroid is dependent on the materials, strengths, size, and relative velocities of the impactor and target.  To complicate matters even more, the size-frequency distribution of the asteroids which produced lunar craters is that of the Earth Crossing Asteroid (ECA) population.  The ECAs are typically derived from Main Belt collision ejecta which are thrown into a secular or mean-motion resonance (\eg Gladman \etal\space 1997) and evolve into Earth-crossing orbits.  These asteroids suffer a different dynamical and collisional evolution and their size-frequency distribution may not reflect that of the Main Belt.  Similarly, determining the size of an asteroid from its absolute magnitude is problematic because they are related by the albedo which is typically unknown.  Muinonen \etal\space (1995) have  discussed the problem of the actual size distribution being sensitive to the relative proportion of asteroids of different albedos in the observed magnitude distribution. 

One of the major problems with analyzing the results of an asteroid survey is compensating for the observational bias.  Magnitude limited surveys find smaller objects when they are closer to the Earth.  For Main Belt asteroids, defined here as those with semi-major axis between 2.0 and 3.5 AU and eccentricity of less than 0.4 (with a perihelion distance of $>$1.3 AU), the distance from the Earth for asteroids at opposition varies from 0.3 to 4.9 AU.  An asteroid with the maximum semi-major axis and eccentricity will exhibit a $\sim$6.5 magnitude change in apparent brightness depending on where it is observed in its orbit.  So the completeness of an asteroid survey is strongly dependent on the limiting magnitude of the system and the size and location of the asteroids.

Biasing effects may be introduced into the data by the observation strategy or methods of asteroid detection.  For instance, asteroids spend less time at perihelion (when they are brighter) than they do at aphelion, and objects on inclined orbits spend less time near the ecliptic than they do at higher latitudes.  Since asteroids move during the exposure time, the magnitude of the asteroid per unit imaging area decreases relative to stars of the same apparent magnitude.  This trailing effect is coupled to the asteroid's orbital elements and the detection efficiency.  Asteroids may move so fast that their trails become too faint for detection.  Those which move too slowly also won't be detected because their motion may not distinguish them during the time interval between exposures.  Between these two limits there exists some efficiency for detecting the asteroid as a function of its rate of motion and brightness.

This paper introduces a new technique for debiasing asteroid survey data in order to obtain the actual number distribution as a function of absolute magnitude, semi-major axis, eccentricity and inclination.  The method is limited only by an understanding of the detector system efficiency and by the computational requirements of the integration.  We present the results of applying this method to data obtained by the Spacewatch telescope.

\section{Early Asteroid Surveys}

The first systematic photographic survey with asteroid magnitudes based on a photometric system was the Yerkes-McDonald Survey (YMS) (Kuiper \etal\space 1958).  In 1950-1952 they twice photographed the area within 20 \deg \space of the ecliptic to a limiting photographic asteroid magnitude of $\sim$16.5.  They found 1550 asteroids and were able to determine the magnitudes for about \twothirds\space of these objects.  Since they managed to completely photograph the entire ecliptic they assumed that the survey was complete to their limiting magnitude.

It was almost another twenty years before the Palomar-Leiden Survey (PLS) (van Houten \etal\space 1970) extended the magnitude-frequency distribution to a photographic magnitude of about 20.  They photographed a 12 \deg$\times$18 \deg \space area of the sky centered on opposition at the vernal equinox and recovered most of those asteroids in fields taken a month later.  Of more than 2000 Main Belt asteroids which they discovered, about 1800 yielded orbits which were sufficiently accurate for them to determine orbital and magnitude frequency distributions. 

The PLS addressed some of the important selection effects in their observational and analysis strategy.  The limited sky coverage of their survey required a large extrapolation from the observed numbers to the total number of asteroids in the Main Belt.  They also applied independent corrections for trailing effects, the inclination bias, and a correction for the number of objects near the detection threshold.  Kres\'{a}k (1971) examined subtle selection effects in the PLS data and showed that orbital perturbations play an important role when interpreting the survey's results.  
   
Extending photographic studies on the scale of the YMS or PLS would be prohibitive.  However, in the past decade, advances in CCD technology as well as computer hardware and software appear to be converging on practical schemes for automated scanning and searching of the entire sky.  Without a doubt, the prototype for these and other CCD scanning sites is the Spacewatch detector system of the University of Arizona, located on Kitt Peak.

\section{The Spacewatch Survey}

The operation of the Spacewatch system has been described in detail elsewhere (Gehrels 1991, Rabinowitz 1991, Scotti 1994, Jedicke 1995).  The analysis in this paper incorporates the same set of data used by Jedicke and Herron (1997) in which 59,226 asteroid observations were found between 23 September 1992 and 8 June 1995.  During this time period, the automated and nearly real-time detection system scanned about 3740 deg$^2$ of the sky to V$\sim$21.  To compare this to the scale of the YMS and PLS, the Spacewatch sky coverage amounts to about half of the area within 10 \deg \space of the ecliptic.  While the Spacewatch program was not designed to determine the magnitude-frequency relation for the Main Belt, it offers a large statistical sample with about one magnitude fainter coverage then the PLS, in a wide area survey, and over a range of absolute magnitudes which overlap both the YMS and PLS.   

Spacewatch observations of Near Earth Asteroids (NEA) were treated by Rabinowitz (1993).  A Monte Carlo model of the NEA orbit population was developed and passed through a simulation of the Spacewatch detector system.  
The Monte Carlo method can be a very effective technique for determining the detector/survey bias.  It may also be computationally expensive since the ensemble of input orbits must be very large in order to reduce the statistical error on the bias calculation to a level comparable to systematic effects.  Care must be exercised in choosing the distribution of `hidden' orbital elements --- those elements for which the detection efficiency is not explicitly calculated.  If the bias is not entirely independent of the hidden elements they may exert undue influence on the calculated detection efficiency for the other elements.  The Monte Carlo technique also fails to provide the analytical basis for the shape of the bias as a function of the various orbit elements.

This paper examines Spacewatch observations of Main Belt asteroids.  We have developed an analytical method which is model independent to compensate for the bias and systematic errors introduced by the observation program and analysis.

\section{Method}

\subsection{Determination of the Detection Bias}

Jedicke (1996) determined the heliocentric probability density ($\rho$) of finding an asteroid with semi-major axis $a$, eccentricity $e$, and inclination $i$ to be
$$
\rho( a, e, i \, ; r,\ell,b) =
     { 1 \over 2 \pi^3 }
     { r \over a } { 1 \over \sqrt{2ar - r^2 + a^2(e^2-1)} }
     { \cos b \over \sqrt { \sin^2 i - \sin^2 b } }
$$
where $r$ is the distance from the Sun, and $\ell$ and $b$ are the heliocentric longitude and latitude respectively.  The derivation of this expression assumes that orbits are randomly distributed in their ascending node, longitude of perihelion, and mean anomaly.  The equation is not applicable to circular and zero inclination orbits.   

The geocentric probability density, as a function of the object's distance from the Earth ($\Delta$), geocentric ecliptic longitude with respect to opposition ($\lambda '$), and latitude ($\beta$), is then given by  
$$
  p( a,e,i \, ; \Delta,\lambda ',\beta) = \rho( a,e,i \, ; r,\ell,b) \biggl | 
     { \partial (r,\ell,b) \over \partial (\Delta, \lambda ', \beta) }
     \biggr |
$$
where the last term in the expression is the Jacobian of the heliocentric$\rightarrow$geocentric co-ordinate transformation.

Assuming that the Earth is on a circular orbit with unit distance from the Sun, aligning the Sun$\rightarrow$Earth vector with the $+x$-axis, and the $+z$-axis with the North Ecliptic Pole, the Jacobian may be determined from these expressions relating the geocentric and heliocentric positions:  
\begin{eqnarray*}
  \Delta \cos\beta \cos\lambda' & = & r \cos b \cos\ell - 1	\\  
  \Delta \cos\beta \sin\lambda' & = & r \cos b \sin\ell	        \\  
  \Delta \sin\beta              & = & r \sin b		            
\end{eqnarray*}

The probability of observing an asteroid with orbital elements $(a,e,i)$ in an area of the sky extending over a range in ecliptic latitude and longitude is then given by
$$
 P( a, e, i ) = \int d\Delta  \int d\lambda'  \int d\beta \;
		   p( a, e, i \, ; \Delta, \lambda', \beta) 
$$ 
The evaluation of this expression is considerably simpler for a survey which scans along great circles in the ecliptic system.  This reason alone may provide enough motivation for future surveys to scan in this manner.  Spacewatch scans are aligned with the equatorial
system and may be of different length in RA ($\alpha$) but the coverage in declination ($\delta$) is fixed by the field of view of the camera.  Therefore, it is more natural to perform the calculation in the equatorial system in determining the Spacewatch bias.
The integration over all Spacewatch scans ($j$) to
obtain the probability of scanning over an asteroid with elements $(a,e,i)$ is given by
\begin{eqnarray*}
 P( a, e, i ) & = &
	\sum_j  \int_0^\infty d\Delta  
                \int_{\alpha_j^{min}}^{\alpha_j^{max}}  d\alpha  
                \int_{\delta_j^{min}}^{\delta_j^{max}}  d\delta  \\
	& \times &    \rho( a, e, i ; r, \ell, b) 
                      \biggl | { \partial (r,\ell,b) \over 
                                 \partial (\Delta, \lambda ', \beta) } \biggr |
                      \biggl | { \partial (\lambda',\beta) \over 
                                 \partial (\alpha,  \delta) } \biggr |
	\epsilon ( a, e, i ; r, \ell, b ;\Delta, \lambda ', \beta, ... )  
\end{eqnarray*}
where the Jacobian of the transformation from geocentric ecliptic to equatorial co-ordinates has also been introduced.  The $min$ and $max$ superscripts on the RA and declination integrations refer to the limits of the Spacewatch scan in each direction.  

The detection efficiency ($\epsilon$) has also been introduced into
this equation.  Its determination as a function of $( a, e, i ; r, \ell, b ;\Delta, \lambda ', \beta )$ is the key component in determining the observational bias since surveys normally maintain excellent records of their scanning location, time and field of view.  

Scanning over the location of an asteroid does not imply that the asteroid was visible or detectable.  A `visible' asteroid is one which has a magnitude above the system's detection threshold.  Depending on the detection technique, it is possible that asteroids which are `visible' may not be detected because of inefficiencies in software or the observer.   A `detectable' asteroid is one which may be found by the techniques used within a group (\eg blinking or automated detection).  Asteroids which move too fast or too slow may not be `detectable' with different techniques and the efficiency for detection may vary with other factors (exposure time, image scale, etc).  

The efficiency of the Spacewatch detector system for finding Main Belt asteroids was discussed in detail by Jedicke and Herron (1997).  They measured the efficiency as a function of an asteroid's apparent magnitude ($V$) and the number of pixels ($n_{pix}$) on the CCD which the asteroid moved between the first and third observation ($\Delta t$).  Limited statistics in the efficiency determination forced them to assume that the two efficiencies were independent of one another such that $\epsilon(V,n_{pix}) = \epsilon(V) \epsilon(n_{pix})$ where $n_{pix}=\omega\Delta t$ and $\omega$ is the asteroid's rate of motion.  

We adopted the IAU's Two-Parameter Magnitude System for Asteroids (Bowell \etal\space 1989) in order to calculate the apparent magnitude $V(H,\Delta,r)$ in the determination of the bias, and the absolute magnitude $H(V,\Delta,r)$ for the data.  In this system, an asteroid's apparent brightness depends upon two parameters intrinsic to the asteroid ($H$, $G$) as well as its geometric relationship with the Sun and Earth.  The geometrical circumstances of an observation are usually expressed in terms of the heliocentric or geocentric distance, and the phase angle subtended between the Earth and Sun as seen from the asteroid.  An asteroid's absolute magnitude is related to its mean area and albedo as viewed from the Earth.  The slope parameter ($G$) is indicative of the manner in which the apparent brightness changes as a function of the phase angle.  We have followed the contemporary practice of fixing the slope parameter at 0.15 in this analysis.  Variations in the apparent magnitude due to changes in aspect or rotation are negligible for the purposes of this work.

The detection efficiency may be coupled to the orbit elements with the use of relations developed by Jedicke (1996).  These provide the four possible motion vectors (or rates of motion $\omega_k$ with $k=1,4$) of an object given its heliocentric location and $(a,e,i)$.

Finally, the probability that the Spacewatch system could detect an asteroid with orbital elements $(a,e,i)$ and absolute magnitude $H$ is given by
\begin{eqnarray*}
 P( a, e, i, H ) & = &
     \sum_j \;  \int_0^\infty d\Delta  \;
                \int_{\alpha_j^{min}}^{\alpha_j^{max}}  d\alpha \; 
                \int_{\delta_j^{min}}^{\delta_j^{max}}  d\delta \;\;
	           \rho( a, e, i \, ; r, \ell, b)                    \\
	&\times&      \biggl | { \partial (r,\ell,b) \over 
                                 \partial (\Delta, \lambda ', \beta) } \biggr |
                   \; \biggl | { \partial (\lambda',\beta) \over 
                                 \partial (\alpha,  \delta) } \biggr | \;
                \epsilon[V(H,r,\Delta)] \;
                \epsilon_{sys} \;
        \sum_k  \epsilon[\omega_k(a,e,i\, ;r,\ell,b) \Delta t_j]
\end{eqnarray*}
The $\epsilon_{sys}$ factor is described below --- it is an efficiency used in compensating for systematic observational and analysis effects.

This unruly expression may be tamed by natural restrictions on the integrals.  For instance, the range of $\Delta$ is restricted in the direction $(\lambda',\beta)$ to a heliocentric distance where $a(1-e) < r < a(1+e)$.  Similar simplifications in the integration may be obtained while integrating over inclination since it is impossible for an asteroid to appear at a heliocentric latitude which is greater than its orbital inclination $(b \le i)$. 
The Jacobian of the equatorial$\rightarrow$ecliptic transformation has a particularly simple form but the geocentric$\rightarrow$heliocentric Jacobian is not so trivial.  In order to increase the execution speed of the code we determined that at a fixed geocentric distance, this Jacobian varied by $<1$\% over the allowed ranges in $(\lambda',\beta)$.   We then applied the value of the Jacobian only as a function of $\Delta$ calculated at opposition.

We implemented the integration for the bias calculation using a variable step size in $\Delta$ since $\rho(a,e,i\,;r,\ell,b)$ is very sensitive to $r(\Delta)$ near perihelion and aphelion.  The integration was tuned to provide about 100 steps for each integration over the allowed geocentric distance.  The field of view of a single Spacewatch CCD frame was about $0\fdg6\!\times\!0\fdg6$ so we chose two steps in $\delta$ and as many
$\sim0\fdg3$ steps in $\alpha$ as were required to cover the scan length (up to 18 CCD frames).  Special consideration was made for any remnant part of a scan after taking an integral number of steps in right ascension.

The grid spacing for the evaluation of the bias ($P$) was motivated by the resolution of the Spacewatch detector in each dimension of $(a,i,H)$ and at reasonable steps in the eccentricity.  The resolution in $(a,i,H)$ was determined by generating two positions with a time separation of one hour for a large sample of asteroid orbits.  These positions were `smeared' on the sky-plane to simulate the $\sim0\farcs6$ absolute astrometric precision of the Spacewatch measurements.  Circular orbits were then calculated from these fake positions allowing the measured inclination, semi-major axis and absolute magnitude to be compared to the known values.  By limiting the study to objects found within 20 \deg \space of opposition in ecliptic longitude we achieved an RMS error of about 0.1 AU in $a$, 2 \deg \space in $i$, and 0.7 magnitudes in $H$.  To complete the integration in a reasonable time and achieve a resolution in $P(a,e,i,H)$ consistent with the detector system, the bias was evaluated in the ranges and steps specified in Table \ref{table:ranges}.

The use of circular orbits is an unfortunate consequence of an observation program geared towards NEAs.  The concomitant loss of eccentricity in the calculation `smears' the values of $(a,i,H)$ and hobbles the implementation of the 4-dimensional model-independent bias calculation.  But the RMS errors are not as bad as might be imagined and the empirical results will show features at the quoted resolution.  This debiasing method cries out for application to a large number, long arc, database of accurate asteroid observations.

\begin{figure}
  \plotfiddle{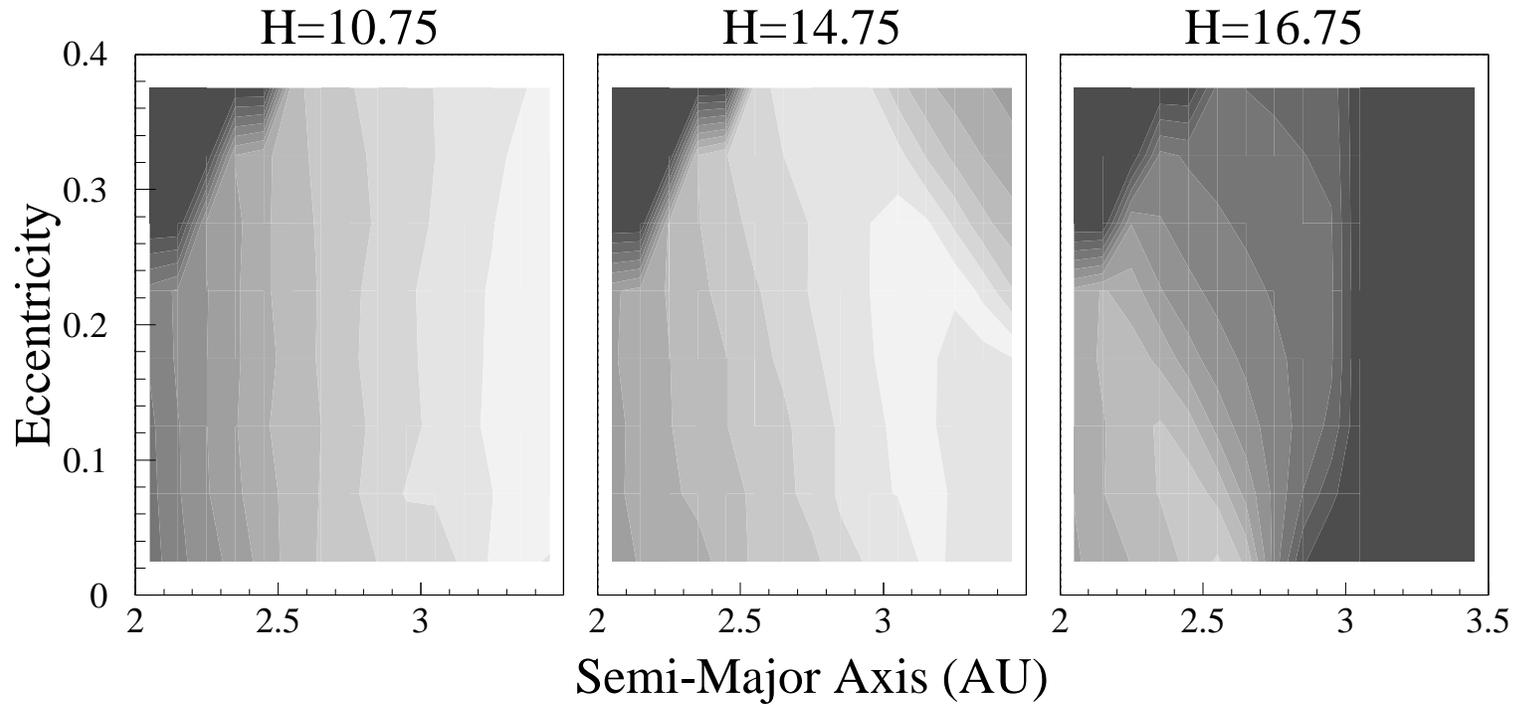}{550pt}{0}{90}{90}{-300pt}{250pt}
  \vskip -1.5in
  \caption{
The probability of asteroid detection (bias or $P$) as a function of semi-major axis ($a$) and eccentricity ($e$) for three discrete values of the absolute magnitude ($H$) and for an inclination of 8 \deg.  The darkest areas correspond to zero detection probability while the brightest areas represent a detection probability of about 7\%.  The region of $P=0$ in the upper left corner corresponds to Mars crossing asteroids which are not included in this study.  The bias calculation was not performed in this region as a time-saving measure.
}
  \label{figure:bias}
\end{figure}

The four-dimensional bias which results from the integration given above yields overall results which correspond well with expectations.  However, the interplay between the various aspects of speed near perihelion, distance at aphelion, detector efficiencies and limiting magnitude, delivered some unexpected results.  Figure \ref{figure:bias} shows three `slices' of the bias as a function of $(a,e)$ for asteroids with an inclination of 8 \deg \space and three values of absolute magnitude.  The probability of detecting large objects ($H=10.75$) which are visible through the entire belt increases with semi-major axis because more distant objects move slower and spend more time in the search volume.  Smaller asteroids ($H=14.75$) in the outer region with low eccentricity have a lesser detection probability by Spacewatch than moderate-$e$ asteroids since they are less likely to be above the limiting magnitude of the system.  Note that the detection probability for these same asteroids decreases when they have a high eccentricity because they spend much of their orbit far from the Sun and Earth.  The final Fig. \ref{figure:bias}C shows the peak of the bias for detection near $(a,e)=(2.5 AU,0.025)$.  The peak value at this location is reduced from that in Figs. \ref{figure:bias}A and B and there is a slow decrease of the bias to high eccentricities and moderate semi-major axis.

It is simple to calculate the expected bias for an asteroid on a circular, 
zero inclination orbit when found at opposition, and to verify that our 
calculation approaches this value as $e,i \rightarrow 0$.  We have also 
compared our calculations for a one degree square at the Vernal Equinox at 
opposition with Spahr's (personal communication, 1997) Monte Carlo simulation 
of an asteroid survey.  The two techniques agree well and reproduce both 
the shape and absolute values of the bias. 

It is clear from the three bias `slices' shown in Fig. \ref{figure:bias} that the correction is a complicated interdependent function of $(a,e,i,H)$.  It would be inappropriate to apply a bias correction which is limited to a subset of these parameters.  Caution must also be exercised in choosing ranges in the parameters which ensure that the bias is never zero in the region of interest.  Furthermore, these bias values assume that the distributions of the other three angular elements of the asteroid orbits are random.  The application of this bias to Spacewatch data is appropriate since these three angular elements are either not measured or not measured well.

The YMS and PLS partitioned the Main Belt into an inner, middle and outer region defined by $2.0<a<2.6$, $2.6<a<3.0$ and $3.0<a<3.5$ respectively.  These divisions are also convenient for our analysis and will allow comparison with the results of the earlier surveys.  Within each region we determined the range in $(e,i,H)$ for which $P(a,e,i,H)$ was everywhere greater than $P_{min}=0.001$.  The maximum detection probability within the ranges and for the step sizes given above is $\sim$0.141 for $a=3.35 AU$, $e=0.025$, $i=3.0 \deg$, and $H=15.25$.  The minimum requirement on the detection probability is almost arbitrary, but was chosen to provide a good range in allowable $(a,e,i,H)$-space, and is also roughly 1\% of the maximum detection probability.

The most restrictive of the parameters is the absolute magnitude.  Once an $(a,H)$ range was specified which satisfied $P \ge P_{min}$ for all ($a,H$), the range of $(e,i)$ was almost unrestricted (within the bounds imposed on the integration given above).  Since this study uses only circular orbits it was not possible to restrict the bias and data in the $e$-dimension.  Fortunately, we will show that the eccentricity was not very restrictive in the bias.  Our technique for compensating for our lack of knowledge of the eccentricity is described in the next section.  Table \ref{table:ranges} provides the final ranges on $(a,e,i,H)$ for this analysis.

The evaluation of a detector system's bias is clearly not trivial.  The process described above provides an accurate measure of the bias given the scanning pattern of the telescope and a detailed knowledge of the operating performance of the detection system (telescope+CCD+software).  Since the observation area is always well known, the most important aspect in calculating the asteroid bias is the detection efficiency.

\subsection{Debiasing the Data} 

\begin{figure}
  \plotfiddle{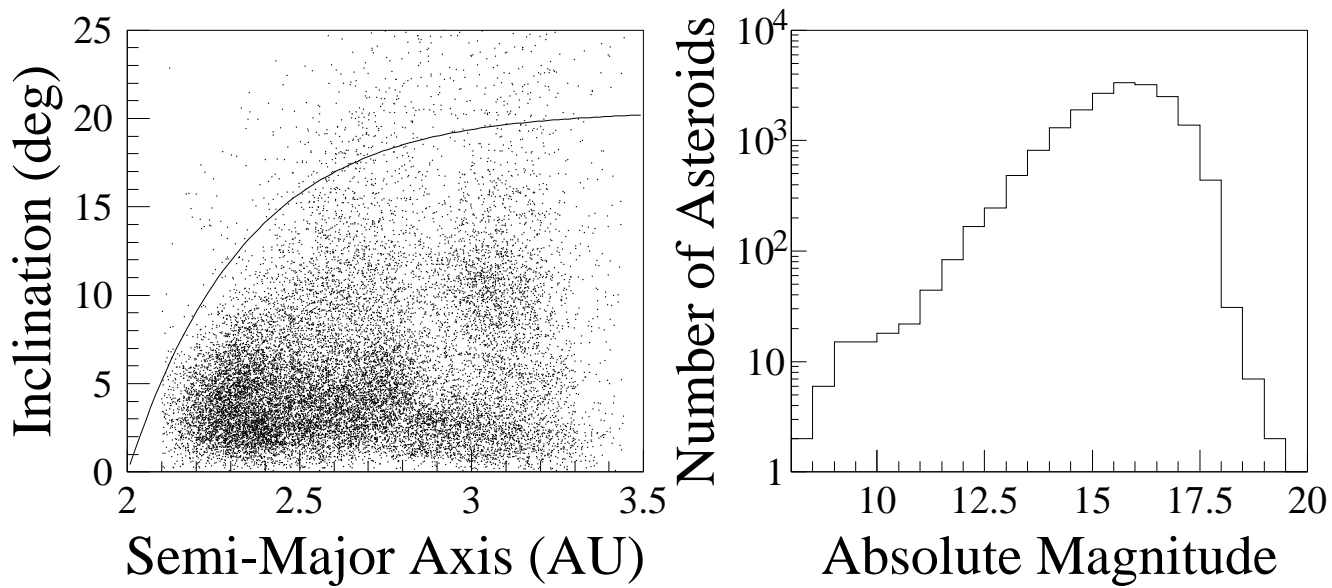}{550pt}{0}{100}{100}{-300pt}{0pt}
  \vskip -1.5in
  \caption{
a) Uncorrected inclination versus semi-major axis for the Main Belt asteroids used in this analysis.  The solid curved line represents the $\nu_6$ resonance.
b) Uncorrected absolute magnitude distribution for the same set of asteroids. 
}
  \label{figure:iah}
\end{figure}

Figure \ref{figure:iah}A shows the $(a,i)$ distribution for 18722 asteroids used in this analysis.  Enhancements in the distribution due to major asteroid families are clear --- especially the Eos group near ($a\sim 3.01 AU$, $i\sim 10 \deg$) --- and the effect of limiting magnitude on the number detected as a function of semi-major axis is obvious.  The secular $\nu_6$ resonance is shown on the Figure as a solid line and its boundary is well represented by the data.  The raw absolute magnitude distribution is shown in Fig. \ref{figure:iah}B.  The limiting magnitude begins to cut into the observed distribution near $H\sim 15$.

The motivation for debiasing is two-fold --- to compensate for the system's detection bias and to extrapolate from the set of observations to the population statistics for the entire Main Belt.  In the previous section we derived an analytic expression for the probability that an asteroid with $(a,e,i,H)$ will be detected by a system.  If we let $d(a,e,i,H)$ represent the detected number density of asteroids, the actual (debiased) number density of asteroids at $(a,e,i,H)$ is given by:
$$
     n(a,e,i,H) = { d(a,e,i,H) \over P(a,e,i,H) }
$$

The absolute number of asteroids ($N$) with semi-major axis in the range 
$a \rightarrow a + \Delta a$, eccentricity in the range 
$e \rightarrow e + \Delta e$, inclination in the range 
$i \rightarrow i + \Delta i$, and absolute magnitude in the range 
$H \rightarrow H + \Delta H$ is obtained by integrating the number density:
\begin{eqnarray*}
  & & N ( a \rightarrow \Delta a,
        e \rightarrow \Delta e,
 	i \rightarrow \Delta i,	
	H \rightarrow \Delta H  )  \\
  &=&  \biggl |  \int_a^{a+\Delta a} da
	\int_e^{e+\Delta e} de
	\int_i^{i+\Delta i} di
	\int_H^{H+\Delta H} dH      { d(a,e,i,H) \over P(a,e,i,H) }
     \biggr |_{P(a,e,i,H) \ne 0}
\end{eqnarray*}
If the probability varies slowly over the specified ranges in the integration variables then it is possible to remove the probability from the integral.  The remaining integral is simply the total number of asteroids ($D$) detected by the system in the same ranges of $(a,e,i,H)$.  Eliminating the notation for the range of elements and magnitude,
$$
    N(a,e,i,H) = { D(a,e,i,H) \over P(a,e,i,H) }
$$
While this expression is exactly what would be expected, the subtlety lies in ensuring that $P(a,e,i,H) \ne 0$.  The PLS and YMS attempted to remove their bias by making relatively harsh cuts on $(a,H)$ in the hope of obtaining a region in $(a,e,i,H)$ in which their bias was almost constant.  Our study imposes cuts on the 4-dimensional $(a,e,i,H)$-space and within this region we have also determined the detailed structure of our bias.

This study is concerned with the one dimensional debiased semi-major axis, inclination and absolute magnitude distributions (represented by $N(a)$, $N(i)$ and $N(H)$ respectively).  These one dimensional distributions are obtained by integrating $N(a,e,i,H)$ over the entire space of the other three variables, subject to the constraint that $P(a,e,i,H) \ne 0$.  \eg 
$$
    N ( H \rightarrow \Delta H )  \\
    =  \Delta H  \biggl |  \int da  \int de  \int di 
                  { d(a,e,i,H) \over P(a,e,i,H) }  \biggr |_{P(a,e,i,H) \ne 0}
$$
In practice, the integrations are coarse summations over $(a,e,i)$ and, neglecting the notation specifying the range in $H$, the numerical integration may be written as
$$
    N ( H ) 
       = \Delta H  \biggl | 
	 \sum_a  \sum_e  \sum_i  { D(a,e,i,H) \over P(a,e,i,H) }
	 \biggr |_{P(a,e,i,H) \ne 0}
$$

A major problem confronting this analysis is that the sets of three Spacewatch observations are typically separated by a total of only about one hour in time.  These short arcs are unsuitable for a determination of all six orbital elements, but our study on the use of circular orbits (described above) shows that they are sufficient to determine $(a,i,H)$ for this analysis.  The use of circular orbits means that we do not have access to the $e$-dimension of the distribution even though we are able to calculate the bias in $e$ for the detector system.

To circumvent our lack of an eccentricity for each asteroid we assumed that the $e$-distribution for all asteroids is semi-independent of $(a,i,H)$.  \ie for some range in $a$, $D(a,e,i,H) = D(a,i,H) D(e)$.  This is certainly not true of the known asteroid belt which is abundant with local enhancements and depletions due to mean motion and secular resonances with the planets.  The disruption of large asteroids due to collisions also produces local enhancements in the orbit element distributions.  However, the low resolution of the Spacewatch detector system in $(a,i,H)$ mutes the effect of assuming that the $e$-distribution is separable.

Proceeding with the assumption that the eccentricity distribution is separable, then
\begin{eqnarray*}
    N ( H ) 
       & = &  \sum_a  \sum_i    D(a,i,H) \; \sum_e  { D(e) \over P(a,e,i,H) } \\
       & = &  \sum_a  \sum_i  { D(a,i,H)  \over P^\prime (a,i,H) }
\end{eqnarray*}
where
$$
  { 1 \over P^\prime (a,i,H) } = \sum_e  { D(e) \over P(a,e,i,H) }.
$$
The bias as a function of $(a,i,H)$ is represented by $P^\prime (a,i,H)$ but the problem remains that the determination of this quantity depends on the observed distribution of objects in eccentricity --- a quantity which can not be measured with the existing Spacewatch data.  

We have verified that the bias of the detector system is only weakly dependent on the eccentricity for a given $(a,i,H)$ so that, for a particular $(a,i,H)$ combination, the {\it detected} $e$-distribution of asteroids will be very similar to the actual population.  Furthermore, this study will show that the set of known asteroids (Bowell \etal, 1994) is probably complete to $H \sim 11$.  Making the assumption that the $e$-distribution for smaller asteroids will be similar to that of the larger (brighter) objects, then $D(a,e,i,H) \sim D(a,i,H) f(e)$.  The term $f(e)$ represents the fraction of the known population of asteroids with $H<11$ with an eccentricity in the range 
$e \rightarrow e + \Delta e$.  In this analysis we considered the eccentricity distributions for the inner, middle and outer regions of the belt separately in the determination of $f(e)$.  The inner and middle regions have similar $e$-distributions while the outer region is distinct with a higher mean eccentricity.  The three dimensional bias
$$
  P^\prime (a,i,H) = \biggl [ \sum_e  { f(e) \over P(a,e,i,H) } \biggr ]^{-1}
$$
may then be used to determine the actual population of asteroids from the observations made by Spacewatch.

We have tested the debiasing method using a Monte Carlo simulation of the Spacewatch detector system (Jedicke and Herron, 1997) and a pathological $(a,e,i,H)$ distribution for the generated Main Belt asteroids.  One of us generated an extremely unrealistic distribution of orbit elements and absolute magnitudes and passed this set of asteroids through the detector simulation to obtain a set of fake `detected' asteroids.  The other author was then able to reproduce the simulated distribution from the `detected' asteroids without the benefit of foreknowledge in what kind of distributions to expect.

\subsection{Systematic corrections and studies}

During the time period of observations used in this study the Spacewatch observing program was usually focused on the discovery of NEAs.  Triplets of images were taken about one-half hour apart in order to detect fast moving asteroids.  This was an effective strategy for detection of the NEAs but it provided only three observations in a 1 to 1.5 hour period --- insufficient for the determination of good orbits and accurate distances for the asteroids in the Main Belt.

Our simulation of Spacewatch observations of Main Belt objects with one-hour arcs fit to circular orbits showed a systematic error in the determination of the semi-major axis.  In the middle of the Main Belt the random errors are about 0.1 AU.  But near the inner edge there is a systematic shift due to it being more likely to find asteroids at perihelion than at aphelion when they are at a heliocentric distance of about 2.05 AU (the center of the first bin in semi-major axis).  An opposite effect occurs at the outer edge of the belt.  The systematic offset was determined as a function of the measured semi-major axis using the simulated data.  Objects with $a$=2.0 AU required a systematic correction of about +0.05 AU and a shift of about -0.03 AU at $a$=3.5 AU.  These corrections are small compared to the bin width of 0.1 AU and, since there are few asteroids near the inner and outer edge of the belt, this correction had little effect on the final debiased $H$ distribution.
 
It was necessary to implement additional factors in the efficiency 
($\epsilon_{sys} = \epsilon_{consistency}(V,\omega) \epsilon_{orbit}$)
to correct for data reduction and quality cuts.  Variable conditions in transparency and seeing, as well as confusion with field stars, made it difficult to determine consistent magnitudes and rates of motion (and therefore orbits) for some of the asteroids.  We required that the observed rate of motion between the first and second observation and then the second and third observation be consistent to within 20\% in right ascension and 30\% in declination, and that the three reported magnitudes have a range of no more than 0.5 magnitudes.  The efficiency of the consistency cut as a function of the asteroid's rate of motion was consistent with being constant ($\sim$90\%) over the range of observed rates ($0\fdg16\>$day$^{-1}<\omega<0\fdg34\>$day$^{-1}$).  The magnitude consistency efficiency was essentially constant ($\sim$85\%) for objects brighter than V$\sim$18 but dropped by about 15\% at the limiting magnitude of V$\sim$21.  Lastly, the determination of a circular orbit did not work for $\sim$5\% of the observation triplets so we used a fixed $\epsilon_{orbit}=0.95$.

Finally, we independently verified the calibration of the Spacewatch V-magnitude system and calculated the effect of CCD saturation on the reported magnitudes.  We found Spacewatch scans which overlapped Landolt's (1992) magnitude calibration regions and compared the magnitudes reported by the software for the Landolt stars to the expected magnitudes.  Since these scans were taken as a matter of course in the overall scanning procedure, they present an unbiased means of determining the magnitude calibration incorporating the averaged effects of airmass, seeing and extinction typically encountered by the program.  We found that the reported V-magnitudes for unsaturated {\it stars} in the data for this analysis had a mean within 0.01 magnitudes of the expected Landolt-V values.  Since the measurement makes no assumption for the point spread function of the object for which the magnitude is being determined, we believe that the asteroid magnitudes are similarly well determined.  Saturation of the CCD pixels began to occur for stars brighter than about $V=13.2$ and we applied the measured stellar saturation correction to asteroids.  Only 0.06\% of the asteroids required this correction which amounted to 0.7 magnitudes at $V=12$.

We have studied the effect of our systematic corrections on each of the results.  All of the corrections (\eg efficiency) vary as a function of a measured quantity (\eg magnitude) so we modified the form of the correction within reasonable limits and re-ran the entire bias calculation and debiasing method.  The checks of the bias calculations were performed on one-fifth of the Spacewatch scans in order to save time.  In particular, we studied the effect of modifying: the efficiency and consistency as a function of both magnitude and motion, the semi-major axis correction, and the eccentricity distributions ($f(e)$).  We also re-ran the bias calculation with steps in $(\alpha,\delta)$ which were one-half and twice the values used in the nominal study.  The step size made almost no difference to the calculated bias values.  The systematic errors presented in the next section are one-half the range in values found in each of the systematic studies.

\section{Absolute Magnitudes and Sizes}

It is tempting to blithely convert absolute magnitudes into diameters ($D$) by assuming an albedo ($p$) for the asteroids --- but this can be dangerous if the results are interpreted too literally since it ignores important selection effects.

For instance, if there are two types of asteroids, one with high albedo and the other with low albedo, there is a stronger bias at a given {\it size} for detecting the intrinsically brighter (high albedo) objects.  Thus, asteroid class ratios at a given size will not be the same as the ratios at a given absolute magnitude.  More rigorously, Harris (personal communication, 1997) has shown that if $R$ is the light:dark albedo ratio ($p_1/p_2$), $a$ is the size population index ($dN/dD \propto D^{-a-1}$), and $f$ is the fraction of objects at constant diameter of high albedo, then the fraction $f^\prime$ of high albedo objects at a given absolute magnitude is
$$
f^\prime = { {f \; R^{a/2}} \over {1 + f \; ( R^{a/2} - 1 )} }.
$$
Rather than using a simple average of the light and dark albedos when converting $H\rightarrow D$, it would be better to use a weighted albedo ($\bar p$) in each region of the belt based on $f^\prime$.  Since this value depends on the albedos, slope parameter and unbiased ratios at a given size, it is clear that the conversion from the absolute magnitude to size realm is fraught with unknowns.

To complicate matters further, the absolute magnitude of an asteroid is not the sole contributor to its `discoverability' --- an asteroid's color can have a serious impact on its detection.  This situation is particularly poignant in comparing the results of the Spacewatch survey to the PLS.  The latter survey used 103a-O film which peaks in sensitivity near 400 nm and cuts off longward of 520 nm.  In contrast, the Spacewatch system begins to have some sensitivity near 300 nm and peaks at about 700 nm.  The peak wavelengths correspond best to the $b$ and $w$ bands (Tholen, 1984) of the Eight-Color Asteroid Survey by Zellner \etal\space (1985).  But the mean $b-v$ and $v-w$ color indices (Tholen, 1984) for C-type asteroids are 0.024 and 0.003 respectively while they are 0.188 and 0.169 for the lighter S-types.  So at any given $V$ ($\sim v$) magnitude (which is used in calculating $H$) Spacewatch is more likely to find S-type asteroids than the PLS.

Gradie \etal\space (1989) provide the bias-corrected distribution of asteroid types in the Main Belt as a function of semi-major axis.  They show that the distribution of taxonomic types is perversely arranged for planetary astronomers --- the darkest asteroids are farthest from the Sun.  If we consider the Main Belt to be composed of light (S-type) and otherwise dark (C, D and P-type) asteroids with albedos of $p_1 \sim 0.155$ and $p_2 \sim 0.05$ respectively (Tholen and Barucci, 1989), we can calculate an appropriately weighted albedo for asteroids in each region.  Using Fig. 3 of Gradie \etal\space (1989) we estimate $f_{inner}=0.5$, $f_{middle}=0.2$ and $f_{outer}=0.07$.  Combining these values with $a=2.5$ for a collisionally evolved equilibrium size distribution (Dohnanyi, 1971), we find $f^\prime_{inner}\sim0.8$, $f^\prime_{middle}\sim0.5$ and $f^\prime_{outer}\sim0.25$.  Finally, at a given $H$, the weighted albedos appropriate for each region of the belt are 
$\bar p_{inner}=0.134$, $\bar p_{middle}=0.103$ and $\bar p_{outer}=0.076$.

We feel that any further attempt at converting our absolute magnitudes to sizes is not justified after careful consideration of these factors.  Sizes for asteroids in each region will be calculated using the weighted albedos, but we caution that these diameters should only be considered as very rough estimates of an asteroid's size.  As a general guide, a ten $km$ diameter asteroid will have an absolute magnitude of about 12.8, 13.1 and 13.5 in the inner, middle and outer regions of the belt respectively.  (A one $km$ diameter asteroid is five magnitudes fainter in $H$.)

\section{Results and Discussion}

\begin{figure}
  \plotfiddle{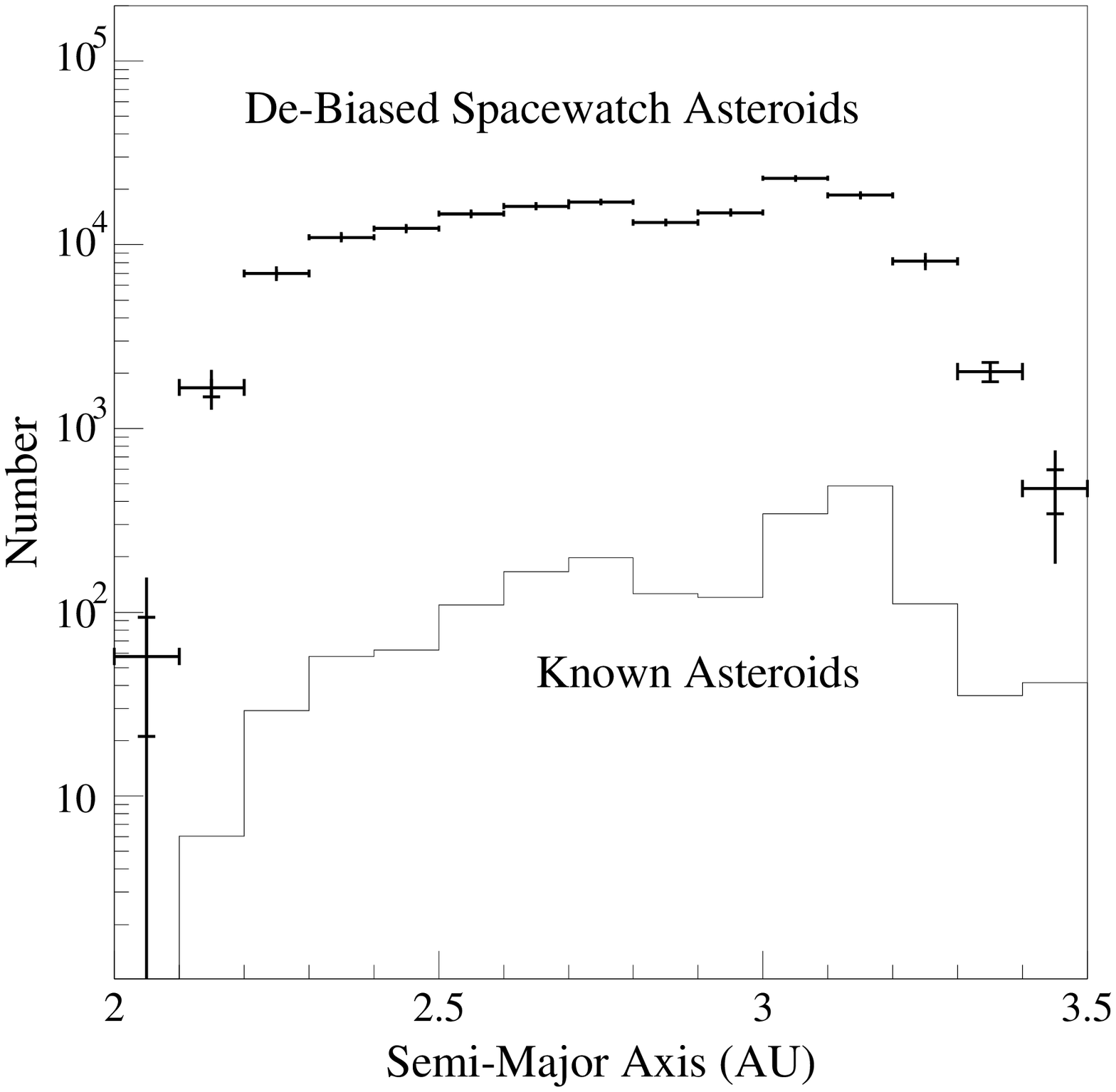}{550pt}{0}{90}{90}{-240pt}{130pt}
  \vskip -1.5in
  \caption{
The bold data points give the semi-major axis distribution of Main Belt asteroids with $11.5 \le H < 16.0$ after correcting for observational bias.  The horizontal extent of each point corresponds to the binwidth.  The small horizontal ticks on the vertical part of the error bar represent the statistical error on the measurement. The total extent of the vertical bar gives the sums of the statistical and systematic errors added in quadrature.  The solid line is the semi-major axis distribution of the known asteroids (Bowell \etal, 1994) with $H<11.5$.
}
  \label{figure:DebiasedSemiaxis}
\end{figure}

\begin{figure}
  \plotfiddle{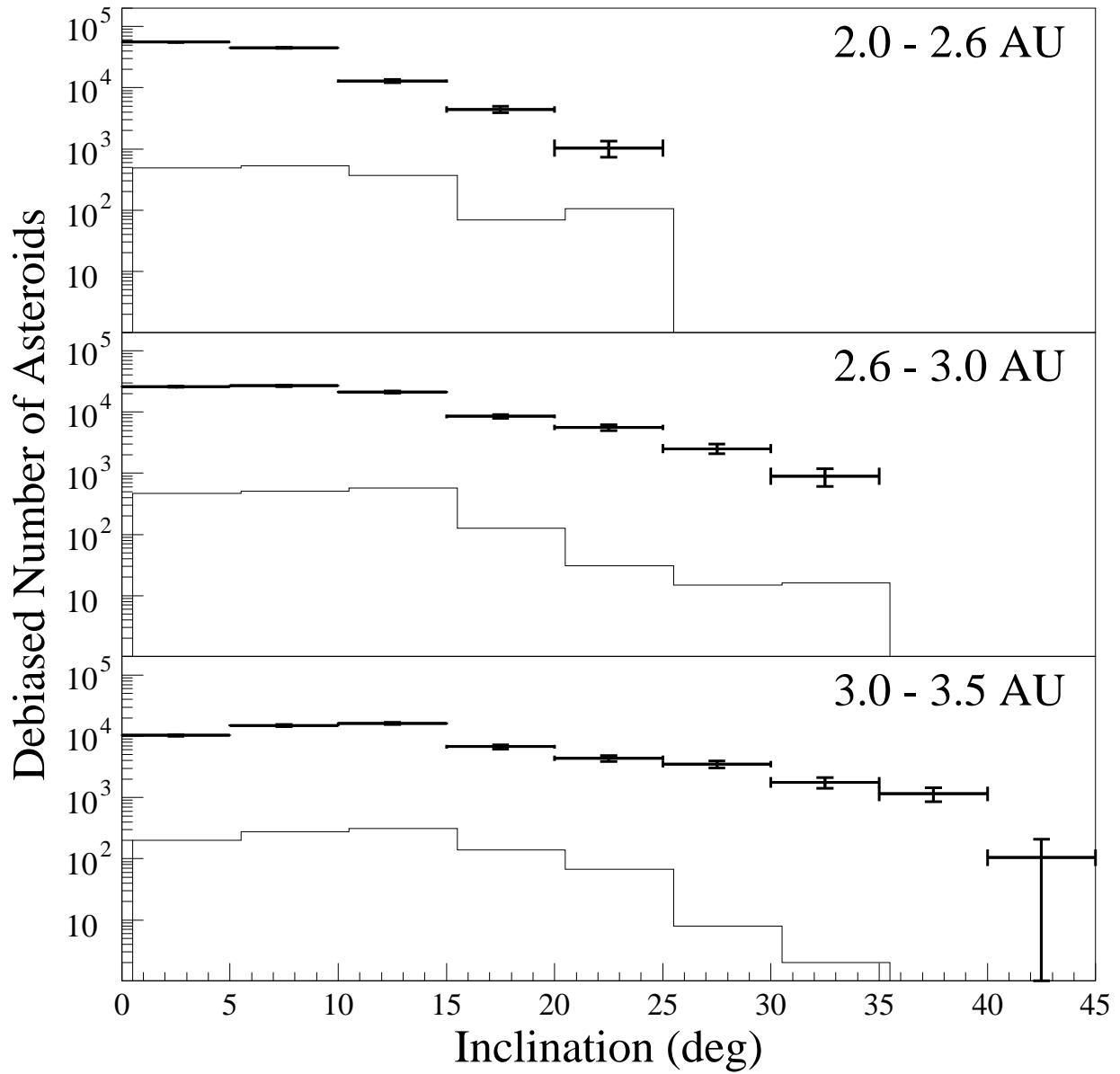}{550pt}{0}{90}{90}{-240pt}{130pt}
  \vskip -1.5in
  \caption{
The bold data points give the inclination distribution of Main Belt asteroids after correcting for observational bias.  The horizontal extent of each point corresponds to the binwidth.  The small horizontal ticks on the vertical part of the error bar represent the statistical error on the measurement.  The total extent of the vertical bar gives the sums of the statistical and systematic errors added in quadrature.  The solid line is the inclination distribution of the known asteroids (Bowell \etal, 1994) to $H=13.0/12.5/11.5$ in the inner/middle/outer regions respectively.  The inner region is truncated for $i>25$ because the Spacewatch bias corrections are excessive for larger inclinations.
}
  \label{figure:DebiasedInclination}
\end{figure}

Figures \ref{figure:DebiasedSemiaxis} and \ref{figure:DebiasedInclination} give the debiased distributions in semi-major axis and inclination for Main Belt asteroids as determined using the technique described above.  The bin size in both dimensions is not fine enough to resolve the mean motion resonances.  Recalling that these distributions were obtained using orbits with arcs of only $\sim$one hour, it is encouraging to see that gross features of the distributions in $a$ and $i$ are visible:  the depletion of asteroids near the inner and outer edges of the belt, the dip near 2.8 AU spanning the range from the 5:2 to the 7:3 mean motion resonance with Jupiter, an increase in the number with heliocentric distance, and the rough slope of the inclination distribution in each region.  The debiased numbers as well as the statistical and systematic errors for the distributions in semi-major axis and inclination are reproduced in Tables \ref{table:semiaxis} and \ref{table:inclination} respectively.

The distribution in semi-major axis of the intrinsically bright (larger) asteroids with $H<11.5$ increases through the belt to a maximum near 3.1 AU.  We will show that the known asteroids are virtually complete to this absolute magnitude so they provide the actual distribution of large asteroids without the confusion of observational bias.  Our technique allows a comparison with the debiased distribution of asteroids in the range $11.5 \le H < 16.0$ (to objects that are roughly an order of magnitude smaller in diameter) throughout the entire belt.  Figure \ref{figure:DebiasedSemiaxis} shows that the semi-major axis distribution of the smaller asteroids is similar to the larger ones.

The peak location and slopes of the inclination distributions of smaller asteroids also mirrors that of the larger asteroids.  The debiased Spacewatch observations show that the peak of the distribution moves to higher inclination for asteroids farther from the Sun.  The strong peak for $10 \deg \le i < 15 \deg$ in the outer region is likely due to the Eos family.  

Note that our observations seem to indicate that the inclination distribution in the outer region extends to considerably higher $i$ than suggested by the set of known asteroids.  Using our simulation of the Spacewatch detector system with circular orbits fit to the `observations' (discussed above) we have verified that these high inclination objects are most likely simply mis-identified lower-$i$ objects.  Normalizing the known data to the Spacewatch debiased data in the third bin, there is an excellent match between the two sets of data for $i<25 \deg$.  In the bin for $30 \deg \le i < 35 \deg$ the debiased distribution is about an order of magnitude higher than the normalized data - exactly the factor predicted by our simulation.

\begin{figure}
  \plotfiddle{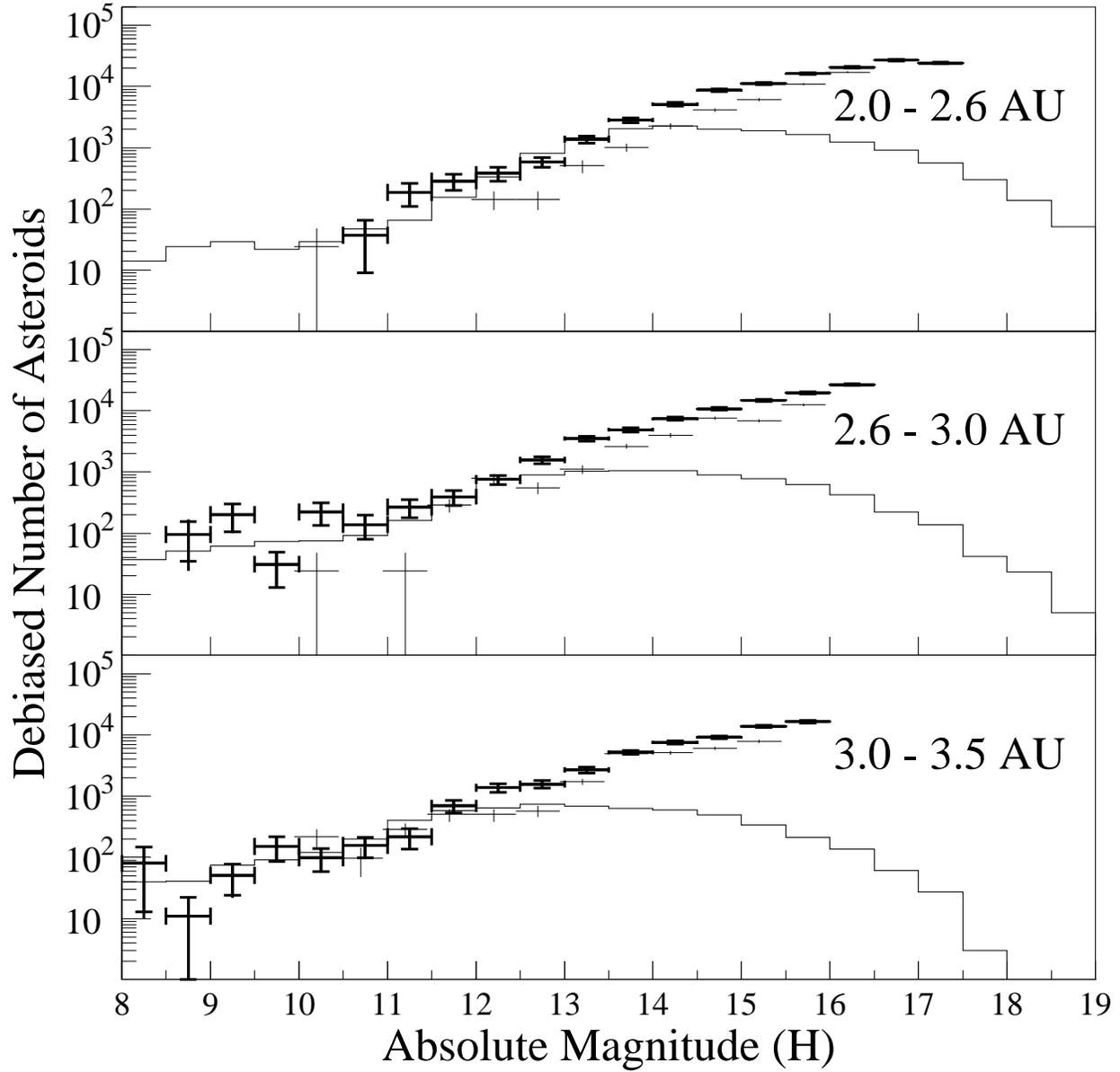}{550pt}{0}{90}{90}{-240pt}{130pt}
  \vskip -1.5in
  \caption{
The thick-lined data points give the debiased number of asteroids in each half-magnitude bin of absolute magnitude.  The thin-lined data points show the debiased results from the PLS (van Houten \etal, 1970) shifted by -0.05 magnitudes in order to reduce confusion with the Spacewatch data.  The solid histogram is the semi-major axis distribution of the known asteroids (Bowell \etal, 1994).
}
  \label{figure:DebiasedAbsmag}
\end{figure}

The main purpose for debiasing the Spacewatch observations was to obtain the corrected absolute magnitude distribution of the Main Belt asteroids shown in Figs. \ref{figure:DebiasedAbsmag}.  The same figure shows the distribution of the known asteroids and also the corrected $H$ distribution determined by the PLS.  Table \ref{table:absmag} provides the debiased number as well as the statistical and systematic errors as a function of absolute magnitude in each region of the belt.

The PLS values were originally calculated in terms of the absolute magnitude $g$ ($=B(1,0)$) and we have used Marsden's (1986) recommendation in converting to $H=g-1$.  Furthermore, the PLS numbers were tabulated (van Houten \etal, 1970, Table 5) in a ``declination strip 18 \deg \space wide'' and require an additional factor of $10^{1.38}$ in order to extrapolate to the entire belt.  Both these factors have been accounted for in Fig. \ref{figure:DebiasedAbsmag}.  The error bars on the PLS data points reflect only the statistical error on the {\it corrected} number of asteroids in their 18 \deg \space wide strip.  The errors are therefore an underestimate of the actual error because we can not account for systematic effects or their correction for loss of detection efficiency in their last bin of photographic magnitudes ($19.5 \le m<20.0$).

Figure \ref{figure:DebiasedAbsmag} shows dramatic differences between the debiased Spacewatch results and the corresponding PLS statistics.  Even after accounting for their correction to the entire belt, the PLS results underestimate the number of {\it known} asteroids in the inner and middle regions of the belt by about a factor of 2 or 3.  The Spacewatch results merge smoothly with the known asteroids and are everywhere above the debiased PLS data.  There are also differences in the slope of the $H$ distribution between the two studies which will be discussed later in more detail.

The PLS debiased values are clearly and systematically low but this may be due to a combination of factors including but not limited to their debiasing technique.  For instance, the relationship $H=g-1$ incorporates both the transformation from photographic to V magnitudes and a difference in the definition of the absolute magnitude functions.  A shift of -0.5 to -1.0 magnitudes for the PLS data in $H$ improves the agreement between their values and both the Spacewatch debiased and known asteroids in the inner and middle regions of the belt.  But applying this same shift to the outer region would worsen their results in comparison to the known asteroids.  

Within the limits of low statistics there is a trend in the ratio of the Spacewatch to PLS debiased data ($R(H)=N_{SW}(H)/n_{PLS}(H)$).  The Spacewatch results are typically a factor of 1.5 to 2 greater than the PLS values, and the ratio decreases from the inner to outer regions.  Even within the regions there is a systematic trend in the relative numbers.  For the inner region, in the range $12<H<16$, the Spacewatch results start at 3 times the PLS values and finish at unity.  The middle region shows no trend in the Spacewatch/PLS ratio and in the outer region the trend is reversed!  In this most distant region of the belt the Spacewatch and PLS results are nearly identical for $H\sim 11$ while the ratio increases to about 1.5 for $H\sim 15$.

Some of the differences between the Spacewatch and PLS results can be understood as a consequence of the taxonomic differentiation of the Main Belt.  In the previous section we discussed how the PLS photographic plates were not sensitive to red light while the Spacewatch system has good efficiency at long wavelengths.  This effect biases the PLS {\it away} from the detection of S-type asteroids which are significantly brighter in that region of the spectrum.  Since the C-type asteroids have relatively flat reflectance spectra, Spacewatch and the PLS are roughly equal in their ability to detect them.  But the Main Belt is strongly skewed in the distribution of taxonomic classes favoring the S-types in the inner region and the C-type and other dark asteroid types in the outer region.  This effect, and the {\it laissez-faire} application of the conversion $H=g-1$ regardless of an asteroid's color, both work towards explaining the discrepancy between the Spacewatch and PLS results.

It is possible that the PLS/SW differences shed light on systematic trends in the actual taxonomic distribution as a function of size (or $H$) within each region of the belt.  If the actual number ratio of S/C types decreases with size in the inner region it would explain why $R(H)$ tends towards unity as $H$ increases in that region.  The systematic trend in $R(H)$ is not as pronounced in the outer region but would still require that the ratio of S/C type asteroids increases with $H$ at larger heliocentric distance.  Taken at face value, the results could imply that the taxonomic differentiation of the belt decreases for smaller asteroids.

Another likely culprit in accounting for the PLS underestimate of the number of asteroids, and also the systematic differences between Spacewatch and the PLS as a function of $H$, lies in their debiasing of the data.  We have shown that there exist intimate links between the bias and $(a,e,i,H)$ while the PLS applied only gross corrections independently in $(a,i,H)$.  Recent calculations of the bias for the PLS (Spahr 1997, personal communication), who used both a Monte Carlo method and the technique described here, show that the PLS survey also suffered from $(a,e,i,H)$-dependent effects.  Failing to account for the fact that the corrections are inter-dependent could have a dramatic effect on the debiased distributions.

Our debiased results clearly show the absolute magnitude to which the set of known asteroids is complete.  We have examined the ratio of the Spacewatch debiased data to the known asteroids and find distinctive breaks from unity at absolute magnitudes near 12.75, 12.25 and 11.25 in the inner, middle and outer regions respectively.  These absolute magnitudes are in excellent agreement with the completeness limit in apparent visual magnitude as determined by Zappal\`{a} and Cellino (1996).  Beyond these limits, the debiased number of asteroids diverges exponentially from the number of known asteroids as a function of their absolute magnitude.  Using the weighted albedos for each region as discussed in the previous section, these absolute magnitudes correspond to diameters of about 10, 15, and 27 km respectively.

Dohnanyi (1971) found that the differential number ($dn$) distribution as a function of radius ($r$) for a self-similar collision cascade in equilibrium is well represented by $dn \propto r^{-a} dr$ with $a=3.5$.  Williams and Wetherill (1994) extended this work and found that the slope of the power-law ($a$) was insensitive to a wide range of factors.  This implies that the number distribution as a function of absolute magnitude for a collisionally evolved set of asteroids should follow a power-law with $\alpha=0.5$ ($N(H) \propto 10^{\alpha H}$).  The two slopes are related by $a=5\alpha + 1$.

It was not possible to fit the debiased Spacewatch results to a single power-law over the measured range of absolute magnitudes in any of the three regions of the Main Belt.  This result extends the conclusion reached by Cellino \etal\space (1991) that Main Belt asteroids larger than about 20-40km diameter can not be fit to a single power-law.  Thus, from the largest asteroids to those only a few km in diameter, there are probably two or three transitions in the slope of the size distribution.  This is most likely due to a size dependent impact-strength since Dohnanyi's (1971) and Williams and Wetherill's (1994) results show that a size-independent impact-strength would yield a single power-law with $\alpha=0.5$ over a wide range of sizes (absolute magnitudes).

We were able to fit most of the debiased results over either two or three intervals in $H$.  The applicable ranges in absolute magnitude and values of the power-law slope are given in Table \ref{table:alpha.global}.  Note that the slope parameter is usually less than the Dohnanyi exponent of 0.5.  A `typical' value in Table \ref{table:alpha.global} is $\alpha\sim 0.3$ which corresponds to $a\sim 2.5$.

Figure \ref{figure:DebiasedAbsmag} and Table \ref{table:alpha.global} indicate that all three regions of the Main Belt show at least one marked change in the slope of the number distribution over the measured ranges in absolute magnitude.  The inner and middle regions each have a slope transition near $H=13$ which corresponds to asteroids about 9 and 11 km in diameter respectively.  Furthermore, the inner region has another slope transition near $H=14.5$ ($\sim$4.5 km) while the outer region has `kinks' near $H=11.5$ ($\sim$25 km) and $H=13.5$ ($\sim$10 km).  In general, there is a trend in which the largest and smallest asteroids in each region have a shallower slope distribution than the medium sized asteroids --- but the meaning of `large' and `small' changes as we move through the regions.  It is also interesting that all three regions of the belt exhibit a transition in slope for asteroids $\sim$10 km in diameter.

\begin{figure}
  \plotfiddle{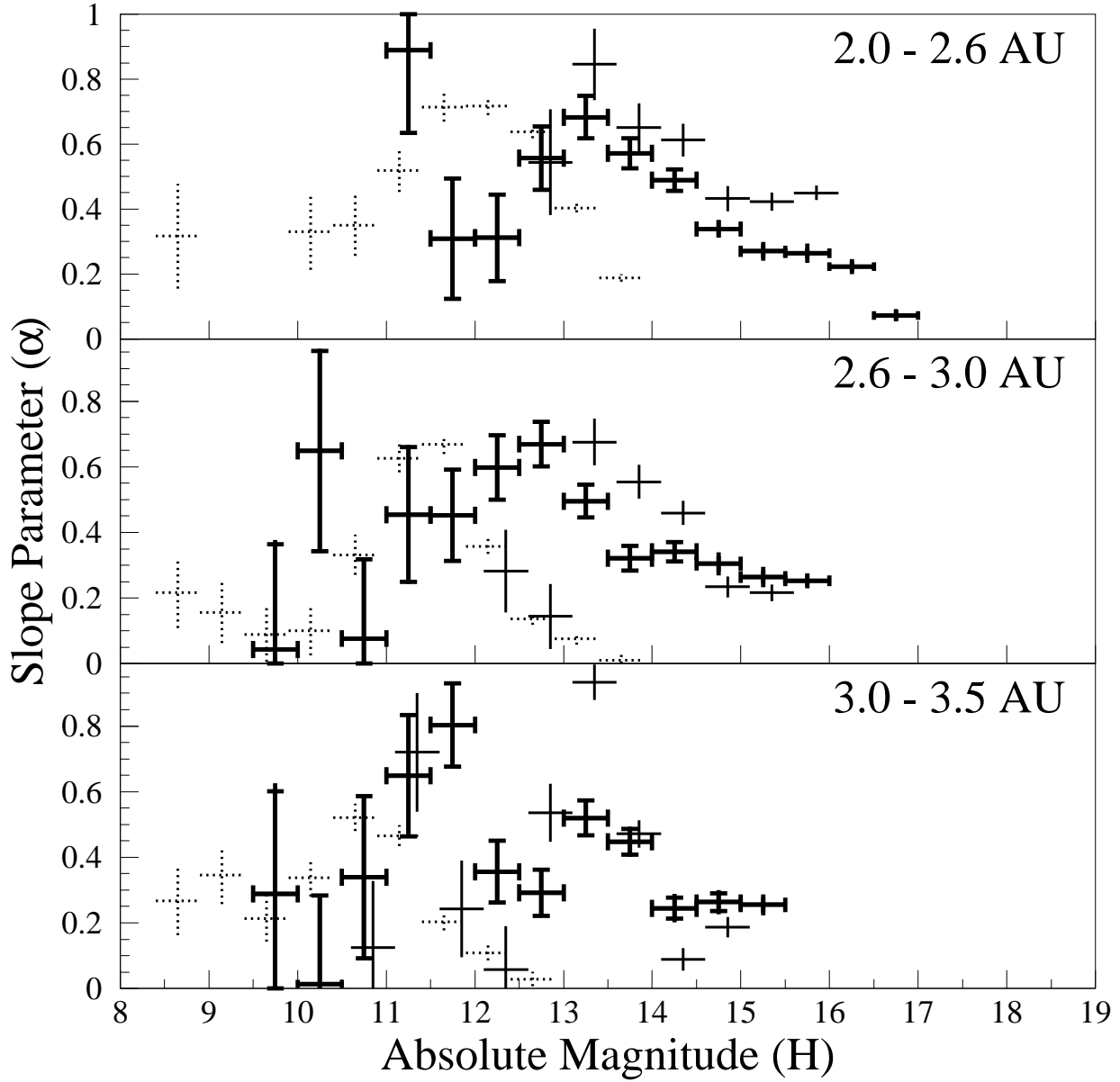}{550pt}{0}{90}{90}{-240pt}{130pt}
  \vskip -1.5in
  \caption{
The slope parameter $\alpha$ ($N(H) \propto 10^{\alpha H}$) as a function of $H$ for three separate regions of the belt.  The bold, normal and dashed data points correspond to this analysis, the PLS (van Houten \etal, 1970) and the set of known asteroids (Bowell \etal, 1994) respectively.  The method of calculating $\alpha$ at each $H$ is described in the text.  For the known asteroids, $\alpha$ is shown whenever $0<\alpha<1$ regardless of the completeness of the data.  The PLS points have been offset by +0.1 magnitudes and the known asteroid points have been offset by -0.1 magnitudes in order to reduce confusion.
}
  \label{figure:alpha}
\end{figure}

To examine the trends in the number distribution in greater detail, Fig. \ref{figure:alpha} shows the {\it instantaneous} value of $\alpha$ as a function of absolute magnitude.  These slopes were obtained by fitting sets of three bins in $\log_{10} N(H)$ to a second-order polynomial from which $\alpha$ and its errors were determined at the central value of $H$.  The values of $\alpha$ are tabulated in Table \ref{table:alpha.local}.

Figure \ref{figure:alpha} exposes the slope transition near $H=13$ in all three regions of the belt as well as the kink near $H=11.75$ in the outer region.  Values for $\alpha$ calculated in the same manner for the PLS and the known set of asteroids are also shown in Fig. \ref{figure:alpha}.  Even though the debiased Spacewatch data differs from the PLS data in both normalization and slope, both data sets show the same kinks and trends in the absolute magnitude distribution.  The point at which the known data become incomplete is clearly visible in each region of the belt.

While the power-law slope meanders around the theoretically expected value of $0.5$ there is much more structure to the $H$-distribution.  Recalling the caveats of the previous section, and understanding that the large error bars on $\alpha$ and $H$ in Fig. \ref{figure:alpha} exacerbate the problem of the conversion from $H$ to $D$, the kinks near absolute magnitudes 13.25, 12.75 and 13.25 in the inner, middle and outer regions correspond to asteroid diameters of about 8.5, 12.5 and 11.0 km respectively.  The kink at $H\sim11.75$ in the outer region represents an asteroid about 22 km in diameter.

Zappal\`{a} and Cellino (1996) suggest that asteroid families will dominate the population at small sizes based on the slope of their absolute magnitude distributions above their completeness limits.  These limits varied from 18-44km diameter depending on the family or region of the belt under consideration.  So their `small' asteroids with steep size distribution correspond to the `large' to `medium' sized asteroids in the Spacewatch data.  It is clear from Figs \ref{figure:DebiasedAbsmag} and \ref{figure:alpha} that extrapolating the steep slopes in this size range to much smaller asteroids will overestimate the calculated number of asteroids in the Main Belt.  Using only the slopes for the smallest asteroids in each region from Table \ref{table:alpha.global}, the total number of asteroids with $10 \leq H \leq 18.35$ (corresponding to asteroids greater than $\sim$1km diameter) is somewhat less than 10$^6$.  Our results imply that the steep slopes of the size distributions for the families does not continue to sizes much less than about ten kilometers diameter.

Considerable theoretical effort has been expended on implications of `bumps' or `kinks' in the main belt size distribution.  The suggestion by Kuiper \etal\space (1958) was that the discontinuities indicate the presence of two populations --- the original accreted planetesimals and fragments from subsequent collisions.  This suggestion was followed by Anders (1965, see his Fig. 4), who concluded that the asteroid belt is not in a highly fragmented state and that the present distribution is only a few collisional time steps removed from the original one.

Contemporary explanations for the bumps include (but are not limited to):  a 
primordial planetesimal size distribution which was not a power-law, a 
transition from gravity to strength dominated collisions, and a `wave'-like 
effect due to a strong cut-off in the size distribution for small asteroids.  
Williams and Wetherill (1994) suggested that the fragmentation process becomes 
mass dependent for objects larger than about 10 km in diameter because 
self-gravitation is not important for smaller objects.  On the other hand, 
hydrodynamical simulations of asteroid collisions by Love and Ahrens (1996) 
predict that the transition from strength to gravity dominated collisions 
occurs at diameters as small as $250\pm150$m for stony bodies.  Our debiased 
observations, and especially the `kink' we have found near $H=13$ ($\sim$10 km 
diameter), may be physical manifestations of one or more of these effects.  

If it represents a transition from strength to gravity dominated collisions as 
suggested by Williams and Wetherill (1994) it would be in conflict with Love 
and Ahrens (1996) more recent calculation.  Our results show no further 
evidence of structure indicative of this transition for objects as small as 
1.5 km, 3.0 km, and 4.5 km diameter in the inner, middle and outer regions 
respectively.  The absolute magnitude distribution must be debiased to $H>22$ 
in order to expose a possible transition for objects $<$0.4 km diameter as 
predicted by Love and Ahrens (1996).

Perhaps the `bump' is the first indication of a `wave' in the size distribution 
induced by a cut-off in the size population for small asteroids.  But the work 
of Davis \etal\space (1993) simulating the cutoff due to radiation effects 
shows a `dip' in the expected size distribution near 10 km diameter.

The fact that the average slope parameter does not asymptotically approach 
Dohnanyi's (1971) value of 0.5 ($a=3.5$) for objects smaller than the `kink' 
may indicate that the size-strength scaling law requires fine-tuning in order 
to match the observed magnitude (size) distribution (Durda and Dermott, 1996).  
The Dohnanyi exponent is applicable only to self-similar collision cascades in 
which the impact strengths are independent of target size.

The actual reason for the `bump' and the varying slope parameter is probably 
some combination of these effects and the relative contribution of each awaits 
a theoretical simulation constrained by these newly debiased observations.

\section{Conclusions}

We have developed a new and analytical technique for calculating the 
observational bias in asteroid surveys.  The method accounts for detection and 
analysis efficiency and may be much faster than Monte Carlo methods.  
Application of the bias calculation and debiasing technique to Main Belt 
asteroids detected by Spacewatch has allowed us to determine their 
distributions in semi-major axis, inclination and absolute magnitude.  The 
debiased distributions in semi-major axis and inclination of the smaller Main 
Belt asteroids mirror that of the brighter (larger) known asteroids.

Based on the deviation of the ratio of the number of known asteroids to 
debiased Spacewatch observations we find that the former is complete to 
absolute magnitudes of about 12.75, 12.25 and 11.25 in the inner, middle and 
outer regions respectively.

We have shown that the number distribution of Main Belt asteroids is not 
consistent with a single power-law in the range $8<H<16$.  It was possible to 
fit a power-law to either two or three intervals in $H$ in each of the three 
regions of the belt.  The slope of the distribution is typically inconsistent 
with the value of 0.5 ($a=3.5$) expected for self-similar collision cascades as 
predicted by Dohnanyi (1971).   This may imply that the impact strength of 
asteroids is size-dependent for objects larger than about 1.5 km, 3.0 km, and 
4.5 km in the inner, middle, and outer regions of the belt.

A distinctive `kink' in the power-law exponent occurs near $H=13$ throughout 
the Main Belt.  Depending on the region, this absolute magnitude corresponds 
to a diameter ranging from about 8.5 to 12.5 km.  The RMS resolution in $H$ of 
$\sim$0.7 magnitudes combined with the lack of an albedo measurement precludes 
an answer to whether the `kink' occurs at constant diameter or constant 
absolute magnitude.

\acknowledgments

We are grateful to the Spacewatch team led by Tom Gehrels for making this work possible.  Special thanks are due to Bill Bottke, Dan Durda, Bob McMillan and Tim Spahr for their support and helpful discussions.  This work is supported by grants from AFOSR, NASA, and private donors.

\references

\reference
  ANDERS, E. 1965.
  Fragmentation History of Asteroids.
  {\it ICARUS} {\bf 4}, 399-408.

\reference
  BELTON, M.J.S., J. VEVERKA, P. THOMAS, P. HELFENSTEIN, D. SIMONELLI, 
      C. CHAPMAN, M.E. DAVIES, R. GREELEY, R. GREENBERG, J. HEAD, S. MURCHIE,
      K. KLAASEN, T.V. JOHNSON, A. McEWEN, D. MORRISON, G. NEUKUM, F. FANALE,
      C. ANGER, M. CARR AND C. PILCHER 1992.
  Galileo Encounter with 951 Gaspra:  First Pictures of an Asteroid.
  {\it SCIENCE} {\bf 257}, 1647-1652.

\reference
  BOWELL, E., B. HAPKE, D. DOMINGUE, K. LUMME, J. PELTONIEMI 
    AND A.W. HARRIS 1989.
  Application of Photometric Models to Asteroids.
  In {\it Asteroids II} (R.P. Binzel, T. Gehrels and M.S. Matthews, Eds.),
  pp. 524-556.  Univ. of Arizona Press, Tucson.

\reference 
  BOWELL, E., K. MUINONEN AND L. H. WASSERMAN 1994.
  A public-domain asteroid orbit database.
  In {\it Asteroids, Comets, Meteors 1993}
  (A. Milani \etal, eds.), pp. 477-481, Kluwer, Dordrecht.

\reference 
  CELLINO, A., V. ZAPPAL\`{A} AND P. FARINELLA 1991.
  The size distribution of main-belt asteroids from IRAS data.
  Mon. Not. R. astr. Soc. (1991) {\bf 253}, 561-574.

\reference
  DAVIS, D.R., P. FARINELLA, P. PAOLICCHI, A.C. BAGATIN, A. CELLINO AND 
    E. ZAPPAL\`{A} 1993.
  Deviations from the Straight Line:  Bumps (and Grinds) in the Collisionally
    Evolved Size Distribution of Asteroids.
  {\it Lunar Planet. Sci. Conf. XXIV}.

\reference
  DOHNANYI, J. S. 1971
  Fragmentation and Distribution of Asteroids.
  In {\it Physical Studies of Minor Planets}
  (T. Gehrels, Ed.), pp. 263-295.
  NASA Scientific and Technical Information Office, Washington, D.C.

\reference
  DURDA, D.D. AND S.F. DERMOTT 1996.
  Size Distributions of Asteroidal Dust:  
    Possible Constraints on Impact Strengths.
  In {\it Physics, Chemistry, and Dynamics of Interplanetary Dust}
  (B.A.S. Gustafson and M.S. Hanner, Eds.),
  ASP Conference Series, {\bf 104}, 473-476. 

\reference
  GLADMAN, B.J., F. MIGLIORINI, A. MORBIDELLI, V. ZAPPAL\`{A}, P. MICHEL, 
    A. CELLINO, C. FROESCHL\'{E}, H.F. LEVISON, M. BAILEY, M. DUNCAN 1997.
  Dynamical Lifetimes of Objects Injected into Asteroid Belt Resonances.
  {\it SCIENCE} {\bf 277}, 197-201.

\reference 
  GRADIE, J.C., C.R. CHAPMAN AND E.F. TEDESCO 1989.
  Distribution of Taxonomic Classes and the Compositional Structure 
    of the Asteroid Belt.
  In {\it Asteroids II} (R.P. Binzel, T. Gehrels and M.S. Matthews, Eds.),
  pp. 316-335.  Univ. of Arizona Press, Tucson.

\reference 
  GEHRELS, T. 1991.
  Scanning with Charge-Coupled Devices.
  {\it Space Science Reviews} {\bf 58}, 347-375.

\reference 
  JEDICKE, R. 1995.
  Automated CCD Scanning for Near Earth Asteroids.
  in {\it New Developments in Array Technology and Applications}
  (A. G. Davis Philip \etal, Eds.), pp. 157-165.

\reference
  JEDICKE, R. 1996.
  Detection of Near Earth Asteroids based upon their Rates of Motion.
  {\it Astron. J.} {\bf 111} (2), 970-982.

\reference
  JEDICKE, R. AND J.D. HERRON 1997.
  Observational Constraints on the Centaur Population.
  {\it ICARUS}, {\bf 127} (2), 494-507.

\reference 
  KRES\'{A}K, L. 1971.
  Orbital Selection Effects in the Palomar-Leiden Asteroid Survey. 
  In {\it Physical Studies of Minor Planets}
  (T. Gehrels, Ed.), pp. 197-210.
  NASA Scientific and Technical Information Office, Washington, D.C.

\reference
  KUIPER, G.P., Y. FUJITA, T. GEHRELS, I. GROENEVELD, J. KENT, 
    G. VAN BIESBROECK AND C.J. VAN HOUTEN 1958.
  Survey of Asteroids.
  {\it Astrophysical Journal, Supplement Series} {\bf 32(III)}, 289-428.

\reference
  LANDOLT, A.U. 1992.
  UBVRI Photometric Standard Stars in the Magnitude Range $11.5-16.0$ 
  Around the Celestial Equator.
  {\it Astron. J.} {\bf 104} 340-371.

\reference
  LOVE, S.G. AND T.J. AHRENS 1996.
  Catastrophic Impacts on Gravity Dominated Asteroids.
  {\it ICARUS} {\bf 124}, 141-155.

\reference
  MARSDEN, B.G., 1986.
  Notes from the IAU General Assembly.  Minor Planet Circ. No. 10193.

\reference
  MUINONEN, K., E. BOWELL AND K. LUMME 1995
  Interrelating Asteroid Size, Albedo and Magnitude Distributions.
  {\it Astron. Astrophys.} {\bf 293}, 948-952.

\reference 
  RABINOWITZ, D.L. 1991.
  Detection of Earth-Approaching Asteroids in Near Real Time.
  {\it Astron. J.} {\bf 101} (4) 1518-1529.

\reference 
  RABINOWITZ, D.L. 1993.
  The Size Distribution of the Earth-Approaching Asteroids.
  {\it Astrophysical J.} {\bf 407} 412-427.

\reference 
  SCOTTI, J. V. 1994.
  Computer Aided Near Earth Object Detection.
  In {\it Asteroids, Comets, Meteors 1993}
  (A. Milani \etal, eds.), pp. 17-30. Kluwer, Dordrecht.

\reference
  THOLEN, D.J. 1984.
  Asteroid Taxonomy from Cluster Analysis of Photometry.
  Ph.D. Thesis, Univ. of Arizona.

\reference
  THOLEN, D.J. AND M.A. BARUCCI 1989.
  Asteroid Taxonomy.
  In {\it Asteroids II} (R.P. Binzel, T. Gehrels and M.S. Matthews, Eds.),
  pp. 298-315.  Univ. of Arizona Press, Tucson.

\reference 
  VAN HOUTEN, C.J., I. VAN HOUTEN-GROENEVELD, P. HERGET AND T. GEHRELS 1970.
  The Palomar-Leiden Survey of Faint Minor Planets.
  {\it Astr. Astrophys. Suppl.} {\bf 2}, 339-448.

\reference 
  WILLIAMS, D. R., AND G. W. WETHERILL 1994.
  Size Distribution of Collisionally Evolved Asteroidal Populations:
  Analytical Solution for Self-Similar Collision Cascades.
  {\it ICARUS} {\bf 107}, 117-128.

\reference 
  ZAPPAL\`{A}, V. AND A. CELLINO 1996.
  Main Belt Asteroids:  Present and Future Inventory.
  In {\it Completing the Inventory of the Solar System}.
  (T.W. Rettig and J.M. Hahn, Eds.),
  ASP Conference Series, {\bf 107}, 29-42. 

\reference 
  ZELLNER, B., D.J. THOLEN AND E.F. TEDESCO 1985.
  The eight-color asteroid survey:  Results for 589 minor planets.
  {\it ICARUS} {\bf 61}, 355-416.


\begin{table}
\tablenum{\uppercase\expandafter{\romannumeral 1}}
\begin{center}
\begin{tabular}{ccccc}

Element  &  Integration  &     Range         &      Range        &     Range     \\[-0.15cm]
         &   Step Size   &   Inner Region    &   Middle Region   &   Outer Region\\
\tableline
    a    &     0.1 AU     &  2.0 - 2.6 AU      &  2.6 - 3.0 AU      &  3.0 - 3.5 AU    \\
    e    &     0.05      &  0.0 - 0.4        &  0.0 - 0.4        &  0.0 - 0.4      \\
    i    &     5 \deg     &  0.5 - 25.5 \deg   &  0.5 - 45.5 \deg   &  0.5 - 45.5 \deg \\
    H    &     0.5       & 10.0 - 17.5       &  8.5 - 16.5       &  8.0 - 16.0     \\
\end{tabular}
\end{center}

\caption{Integration step sizes and ranges for orbital elements and 
	 absolute magnitude.}

\label{table:ranges}

\tablecomments{The step sizes in the integration correspond roughly to the resolution of
	 the detector system and the analysis.  The ranges of orbital elements and 
	 absolute magnitude are those for which the
	 Main Belt asteroid detection probability is greater than 0.1\%.}

\end{table}

\clearpage

\begin{table}
\tablenum{\uppercase\expandafter{\romannumeral 2}}
\begin{center}
\begin{tabular}{cccc}

Semi-Major Axis &       Number       &          Number           &           Number          \\[-0.15cm]
    (AU)      &     Inner Region     &        Middle Region      &         Outer Region      \\[-0.15cm]
              &($10.5 \le H < 17.5$) &  ($ 8.5 \le H < 16.5 $)   &   ($ 8.0 \le H < 16.0 $)  \\
\tableline
  2.05  & $   100 \pm  40 \pm  180 $ &                           &                           \\
  2.15  & $  4700 \pm 300 \pm 1100 $ &                           &                           \\
  2.25  & $ 19000 \pm 600 \pm 1000 $ &                           &                           \\
  2.35  & $ 32200 \pm 900 \pm  700 $ &                           &                           \\
  2.45  & $ 30900 \pm 900 \pm  600 $ &                           &                           \\
  2.55  & $ 31500 \pm 900 \pm  600 $ &                           &                           \\
  2.65  &                            & $ 24200 \pm 800 \pm 700 $ &                           \\
  2.75  &                            & $ 25900 \pm 800 \pm 500 $ &                           \\
  2.85  &                            & $ 20000 \pm 800 \pm 700 $ &                           \\
  2.95  &                            & $ 21200 \pm 800 \pm 900 $ &                           \\
  3.05  &                            &                           & $ 24300 \pm 800 \pm 900 $ \\
  3.15  &                            &                           & $ 20900 \pm 800 \pm 600 $ \\
  3.25  &                            &                           & $  9900 \pm 600 \pm 900 $ \\
  3.35  &                            &                           & $  3400 \pm 400 \pm 200 $ \\
  3.45  &                            &                           & $   700 \pm 200 \pm 400 $ \\

\end{tabular}
\end{center}

\caption{Debiased number of asteroids as a function of semi-major axis.}

\label{table:semiaxis}

\tablecomments{The central
 	 value is given for each 0.1 AU bin in $a$.  The table has been split into
         the three regions of the belt to emphasize that the debiased
	 numbers correspond to different ranges in the absolute magnitude.  
	 This table does {\it not} correspond to Fig. \ref{figure:DebiasedSemiaxis}
	 which shows the debiased semi-major axis distribution for the same range
	 of absolute magnitudes throughout the belt. The first error is
	 statistical and the second error is the systematic error as
	 described in the text.}

\end{table}

\begin{table}
\tablenum{\uppercase\expandafter{\romannumeral 3}}
\begin{center}
\begin{tabular}{cccc}

Inclination( \deg)   &    Number        &            Number           &             Number            \\[-0.15cm]
                    &   Inner Region   &          Middle Region      &           Outer Region        \\
\tableline
   3.0  &  $ 55400 \pm  800 \pm 1000 $ &  $ 25700 \pm  500 \pm  500 $ &  $ 10400 \pm  300 \pm  300 $ \\
   8.0  &  $ 44900 \pm 1100 \pm  900 $ &  $ 26900 \pm  800 \pm  600 $ &  $ 15000 \pm  500 \pm  400 $ \\
  13.0  &  $ 12800 \pm  700 \pm  300 $ &  $ 21100 \pm  900 \pm  800 $ &  $ 16200 \pm  700 \pm  500 $ \\
  18.0  &  $  4400 \pm  500 \pm  100 $ &  $  8500 \pm  600 \pm  300 $ &  $  6800 \pm  500 \pm  200 $ \\
  23.0  &  $  1000 \pm  300 \pm   20 $ &  $  5600 \pm  600 \pm  200 $ &  $  4400 \pm  500 \pm  100 $ \\
  28.0  &                              &  $  2500 \pm  500 \pm   70 $ &  $  3500 \pm  500 \pm   70 $ \\
  33.0  &                              &  $   900 \pm  300 \pm   40 $ &  $  1800 \pm  400 \pm   70 $ \\
  38.0  &                              &                              &  $  1100 \pm  300 \pm   20 $ \\
  43.0  &                              &                              &  $   100 \pm  100 \pm    5 $ \\
\end{tabular}
\end{center}

\caption{Debiased number of asteroids as a function of inclination.}

\label{table:inclination}

\tablecomments{The central value is given for each 5 \deg \space
         interval in the inclination.  The errors are
	 statistical and systematic respectively.}

\end{table}

\begin{table}
\tablenum{\uppercase\expandafter{\romannumeral 4}}
\begin{center}
\begin{tabular}{cccc}

Absolute Magnitude  &    Number        &            Number            &             Number         \\[-0.15cm]
                    &  Inner Region    &          Middle Region       &           Outer Region     \\
\tableline
   8.25  &                             &                              & $     80 \pm  70 \pm  20 $ \\
   8.75  &                             & $    100 \pm   60 \pm  40  $ & $     10 \pm  10 \pm   1 $ \\
   9.25  &                             & $    200 \pm  100 \pm  40  $ & $     50 \pm  30 \pm  10 $ \\
   9.75  &                             & $     30 \pm   20 \pm   1  $ & $    150 \pm  70 \pm  10 $ \\
  10.25  &                             & $    220 \pm   90 \pm  20  $ & $    100 \pm  40 \pm  10 $ \\
  10.75  & $     40 \pm  30 \pm   1  $ & $    140 \pm   60 \pm  20  $ & $    160 \pm  60 \pm  20 $ \\
  11.25  & $    190 \pm  80 \pm  10  $ & $    270 \pm   90 \pm  20  $ & $    220 \pm  80 \pm  20 $ \\
  11.75  & $    290 \pm  90 \pm  20  $ & $    400 \pm  110 \pm  30  $ & $    700 \pm 160 \pm  60 $ \\
  12.25  & $    380 \pm 100 \pm  30  $ & $    750 \pm  130 \pm  50  $ & $   1400 \pm 200 \pm 100 $ \\
  12.75  & $    590 \pm 100 \pm  50  $ & $   1600 \pm  200 \pm 100  $ & $   1600 \pm 210 \pm 120 $ \\
  13.25  & $   1400 \pm 180 \pm 100  $ & $   3500 \pm  330 \pm 250  $ & $   2700 \pm 280 \pm 180 $ \\
  13.75  & $   2800 \pm 250 \pm 200  $ & $   4900 \pm  350 \pm 330  $ & $   5200 \pm 400 \pm 300 $ \\
  14.25  & $   5100 \pm 300 \pm 400  $ & $   7400 \pm  400 \pm 400  $ & $   7500 \pm 450 \pm 300 $ \\
  14.75  & $   8700 \pm 400 \pm 550  $ & $  10700 \pm  500 \pm 400  $ & $   9200 \pm 500 \pm 330 $ \\
  15.25  & $  11200 \pm 500 \pm 500  $ & $  14900 \pm  600 \pm 300  $ & $  13800 \pm 600 \pm 550 $ \\
  15.75  & $  16200 \pm 600 \pm 300  $ & $  19600 \pm  700 \pm 700  $ & $  16500 \pm 800 \pm 900 $ \\
  16.25  & $  20500 \pm 600 \pm 500  $ & $  26600 \pm 1000 \pm 1300 $ &                            \\
  16.75  & $  27000 \pm 800 \pm 1000 $ &                              &                            \\
  17.25  & $  24200 \pm 900 \pm 1100 $ &                              &                            \\
\end{tabular}
\end{center}

\caption{Debiased number of asteroids as a function of absolute magnitude.}

\label{table:absmag}

\tablecomments{The central
 	 value is given for each 0.5 magnitude wide bin in the absolute magnitude.
	 The errors are statistical and systematic respectively.}

\end{table}

\begin{table}
\tablenum{\uppercase\expandafter{\romannumeral 5}}
\begin{center}
\begin{tabular}{cccc}

 Region &     Absolute Magnitude    & $\alpha$   &     $C$   \\[-0.15cm]
        &           Range           &                        \\

\tableline

        &   $ 11.0 \le H < 13.0 $   &  $0.323$   &  $-1.38$  \\
 Inner  &   $ 13.0 \le H < 14.5 $   &  $0.558$   &  $-4.28$  \\
        &   $ 14.5 \le H < 17.0 $   &  $0.248$   &  $0.290$  \\

\tableline

 Middle &   $ 10.5 \le H < 13.0 $   &  $0.536$   &  $-3.66$  \\
        &   $ 13.0 \le H < 16.5 $   &  $0.287$   &  $-0.228$ \\

\tableline

        &   $ 8.0 \le H < 11.5 $    &  $0.245$   &  $-0.435$ \\
 Outer  &   $ 11.5 \le H < 13.5 $   &  $0.350$   &  $-1.21$  \\
        &   $ 13.5 \le H < 16.0 $   &  $0.250$   &  $0.298$  \\

\end{tabular}
\end{center}

\caption{Parameterization of the number distribution as a function of absolute magnitude.}

\label{table:alpha.global}

\tablecomments{The debiased number ($N$) of asteroids in the specified range and
	 per half-magnitude in $H$ in each region is well represented by
	 $\log_{10} N = \alpha H + C$.  This representation is
	 not continuous at the transitions in $H$.}

\end{table}

\begin{table}
\tablenum{\uppercase\expandafter{\romannumeral 6}}
\begin{center}
\begin{tabular}{cccc}

Absolute Magnitude  &  Inner Region   &         Middle Region      &          Outer Region      \\[-0.15cm]
                    &   $\alpha$      &          $\alpha$          &           $\alpha$         \\
\tableline
   8.75  &                            &                            & $-0.20 \pm 0.53 \pm 0.17 $ \\
   9.25  &                            & $-0.49 \pm 0.41 \pm 0.17 $ & $ 1.14 \pm 0.41 \pm 0.04 $ \\
   9.75  &                            & $ 0.04 \pm 0.32 \pm 0.09 $ & $ 0.29 \pm 0.31 \pm 0.13 $ \\
  10.25  &                            & $ 0.65 \pm 0.31 \pm 0.06 $ & $ 0.01 \pm 0.27 \pm 0.05 $ \\
  10.75  &                            & $ 0.08 \pm 0.24 \pm 0.02 $ & $ 0.34 \pm 0.25 \pm 0.02 $ \\
  11.25  & $ 0.89 \pm 0.25 \pm 0.02 $ & $ 0.45 \pm 0.21 \pm 0.05 $ & $ 0.65 \pm 0.18 \pm 0.03 $ \\
  11.75  & $ 0.31 \pm 0.19 \pm 0.03 $ & $ 0.45 \pm 0.14 \pm 0.02 $ & $ 0.80 \pm 0.13 \pm 0.02 $ \\
  12.25  & $ 0.31 \pm 0.13 \pm 0.02 $ & $ 0.60 \pm 0.10 \pm 0.02 $ & $ 0.36 \pm 0.09 \pm 0.01 $ \\
  12.75  & $ 0.56 \pm 0.10 \pm 0.03 $ & $ 0.67 \pm 0.07 \pm 0.02 $ & $ 0.29 \pm 0.07 \pm 0.01 $ \\
  13.25  & $ 0.68 \pm 0.07 \pm 0.02 $ & $ 0.50 \pm 0.05 \pm 0.01 $ & $ 0.52 \pm 0.05 \pm 0.01 $ \\
  13.75  & $ 0.57 \pm 0.05 \pm 0.01 $ & $ 0.32 \pm 0.04 \pm 0.01 $ & $ 0.45 \pm 0.04 \pm 0.02 $ \\
  14.25  & $ 0.49 \pm 0.03 \pm 0.01 $ & $ 0.34 \pm 0.03 \pm 0.02 $ & $ 0.24 \pm 0.03 \pm 0.03 $ \\
  14.75  & $ 0.34 \pm 0.03 \pm 0.01 $ & $ 0.30 \pm 0.02 \pm 0.03 $ & $ 0.26 \pm 0.03 \pm 0.03 $ \\
  15.25  & $ 0.27 \pm 0.02 \pm 0.02 $ & $ 0.26 \pm 0.02 \pm 0.02 $ & $ 0.25 \pm 0.02 \pm 0.02 $ \\
  15.75  & $ 0.26 \pm 0.02 \pm 0.02 $ & $ 0.25 \pm 0.02 \pm 0.01 $ &                            \\
  16.25  & $ 0.22 \pm 0.01 \pm 0.02 $ &                            &                            \\
  16.75  & $ 0.07 \pm 0.01 \pm 0.01 $ &                            &                            \\
\end{tabular}
\end{center}

\caption{The slope parameter $\alpha$ as a function of absolute magnitude.}

\label{table:alpha.local}

\tablecomments{The central value is given for each 0.5 magnitude bin in $H$. 
	 The first error is statistical and the second is systematic.}

\end{table}

\end{document}